\documentclass[11pt,a4paper]{article}
\usepackage[hyperref]{emnlp2018}
\usepackage{times}
\usepackage{latexsym}
\usepackage{amsmath}
\usepackage{url}
\usepackage{graphicx}
\usepackage{caption}
\usepackage{enumitem}
\usepackage{placeins}
\usepackage{float}

\aclfinalcopy 

\title{A Deep Neural Network Sentence Level Classification Method with Context Information}

\author{Xingyi Song \and Johann Petrak\\
  Department of Computer Science \\
  University of Sheffield \\
  Sheffield, UK \\
  {\tt \{x.song, johann.petrak\}@sheffield.ac.uk} \\\And
  Angus Roberts  \\
  NIHR Biomedical Research Centre \\
  Institute of Psychiatry, Psychology and Neuroscience \\
  King’s College London \\
  London, UK \\
  {\tt angus.roberts@kcl.ac.uk} \\}

\date{}

\begin{document}
\maketitle
\begin{abstract}
  In the sentence classification task, context formed from sentences
  adjacent to the sentence being classified can provide important
  information for classification. This context is, however, often
  ignored. Where methods do make use of context, only small amounts are
  considered, making it difficult to scale. We present a
  new method for sentence classification, Context-LSTM-CNN, that makes use of
  potentially large contexts. The method also utilizes long-range dependencies
  within the sentence being classified, using an LSTM, and short-span features, using a stacked
  CNN.  Our experiments demonstrate that this approach consistently improves
  over previous methods on two different datasets.
\end{abstract}

\section{Introduction}

Artificial neural networks (ANN) and especially Deep Neural Networks (DNN)
give state-of-the art results for sentence classification tasks.  Usually,
sentences are treated as separate instances for the task. However, in many
situations the sentence that is the focus of classification appears in a
context that can provide additional information. For example, in the below sentences from the IEMOCAP dataset, it is difficult to classify M02 as showing excitement, without the prior context:
\begin{itemize}
[topsep=0pt,itemsep=-1ex,partopsep=1ex,parsep=1ex]
\item {\small M01: I got it.  I got accepted to U.S.C..}
\item {\small F01: Oh, for real?}
\item {\small M02: Yes! I just found out today.  I just got the letter.}
\end{itemize}

Our work is motivated by sentence classification in the text of medical records, in which complex judgements may be made across
several sentences, each adding weight and nuance to a point. We
believe, however, that the techniqe is more widely applicable. In
order to test generalisability and to allow reproducibility, we
therefore present an evaluation of the method with publicy
available, non-medical corpora.

Previous work on using context for sentence classification used LSTM and CNN
network layers to encode the surrounding context, giving an improvement in
classification accuracy \cite{Lee2016}. However, the use of CNN and LSTM
layers imposes a significant computational cost when training
the network, especially if the size of the context is large. For this reason,
the approach presented in \cite{Lee2016} is explicitly
intended for sequential, short-text classification.


In many cases, however, the context available is of significant size.  
We therefore introduce a new method, {\bf Context-LSTM-CNN}\footnote{The code is publicly available at https://github.com/deansong/contextLSTMCNN}, which is based on the
computationally efficient FOFE (Fixed Size Ordinally Forgetting)
method \cite{Zhang2015}, and an architecture that combines an LSTM and CNN for
the focus sentence. The method consistently improves over results obtained
from either LSTM alone, CNN alone, or these two combined, with little increase
in training time.

This paper makes three contributions: 1) a demonstration of the importance of
context in some sentence classification tasks;
2) an adaptation of existing datasets for such sentence
classification tasks, in order to support reproducibility of evaluations;
3) a neural architecture for sentence classification that outperforms previous
methods, and can include context
of arbitrary size without incurring a large computational cost.

\section{Related work}

%

Since their introduction \cite{Collobert2011}, CNNs with word
embedding language models have become common for text classification tasks \cite{Kim2014,Conneau2017}.
One limitation of the original CNN approach is the loss of long distance dependencies. 
In order to deal with this in image and speech recognition tasks,
\newcite{Xu2015,Sainath2015} combined CNNs with a Recurrent Neural
Network (RNN) layer. \newcite{Zhou2015} subsequently applied this to text classification.
However, the CNN-RNN approach was originally devised for sequence
labelling, is biased towards later words in the sequence,
and does not perform better than CNN alone. \newcite{Huynh2016} suggested reversing the architecture to first 
apply the RNN followed by a CNN with pooling to obtain global features. This gave results that improved 
over CNN-RNN, but not over CNN alone. 
In this paper, we build on \newcite{Huynh2016}'s approach by replacing the GRU-based RNN \cite{Cho2014} with an LSTM \cite{Hochreiter1997} and 
by using multiple kernel 
sizes and more features in the subsequent CNN layer.

\newcite{Lee2016} showed that when classifying short texts, accuracy can be
boosted by adding a CNN or LSTM derived vector representation of the
surrounding context. For long contexts (such as patient records which
may include well over 100 sentences), however, this will incur a significant additional
computational cost. In this paper, we therefore apply an adaptation of the
FOFE encoding \cite{Zhang2015} to encode context.

\section{Model}

\begin{figure}
\includegraphics[width=0.5\textwidth]{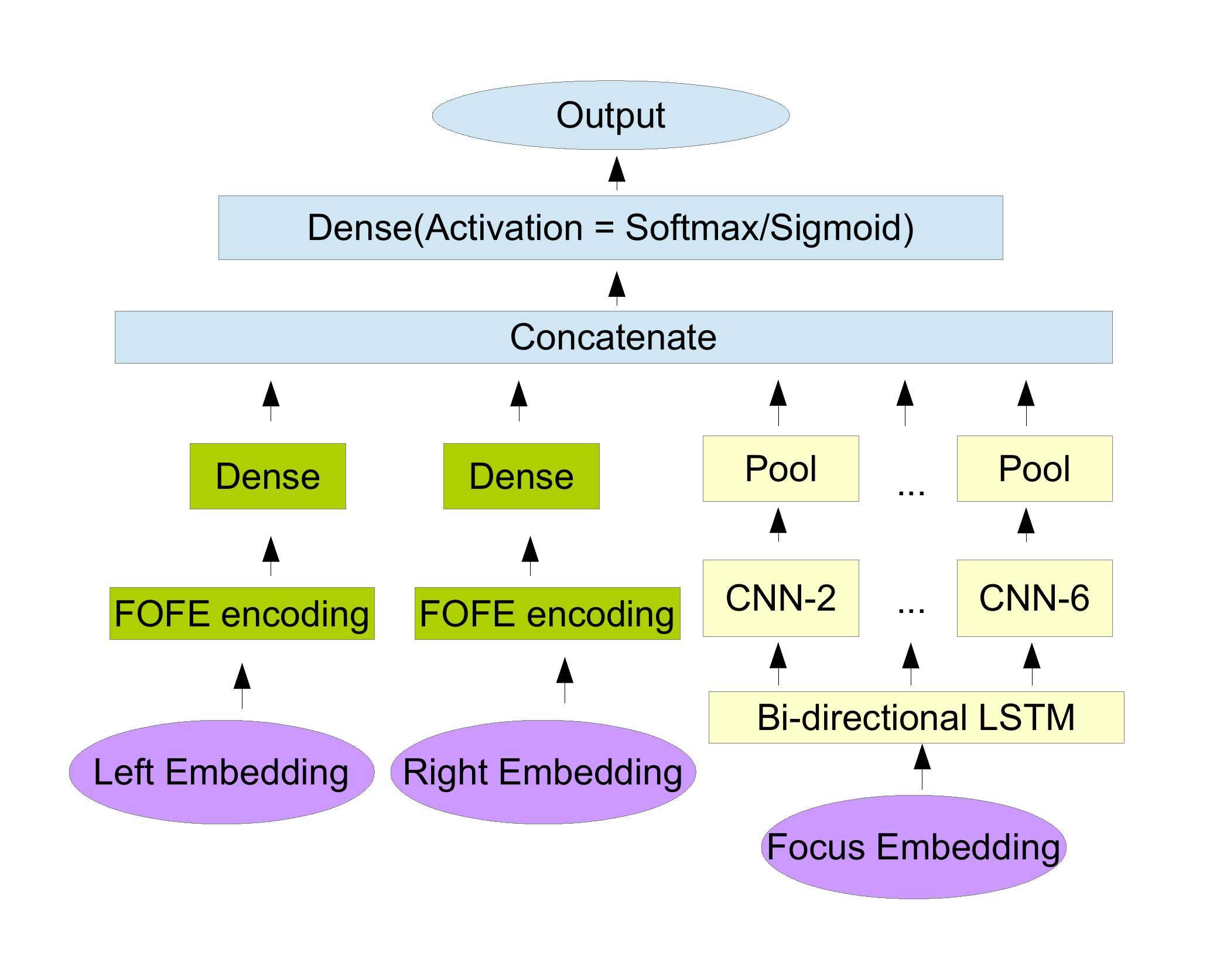}
\caption{Structure of the C-LSTM-CNN model}\label{figmodel}
\end{figure}

The Context-LSTM-CNN model is shown in Figure \ref{figmodel}. It is based on the following components:

\begin{enumerate}[topsep=0pt,itemsep=-1ex,partopsep=1ex,parsep=1ex]
\item Input layer using word embeddings to encode the words of the focus sentence.
\item Bi-directional LSTM applied to the word embeddings of the focus sentence.
\item CNN on the outputs of the LSTM.
\item FOFE applied to word embeddings of both left and right context. 
\item A final output layer.
\end{enumerate}

In brief, an LSTM
layer is used to encode the focus sentence. This is followed by convolutional
layers with small-size kernels and max-pooling to extract local features at
specific points from the LSTM outputs. In addition to processing the focus sentence, we also encode the full
left and right contexts using an adaptation of FOFE applied to our embeddings. 
This encodes any variable length context into a fixed length
embedding, thus allowing us to include large contexts without rapidly
increasing the computational cost. The output of the FOFE
layers are then each passed through separate fully connected layers, before
being concatenated and connected to  output layer.

In detail, the full network takes three inputs. The first is the sequence of words
$X = (x_1,x_2,...x_T)$, where $T$ is the length of the sentence to be
classified, and where each $x_i$ is a word embedding for the respective word
in this sentence. Embeddings are pre-trained by Word2Vec
\cite{NIPS2013_5021} on the corpus used for the respective experiment. The embeddings are not updated during the training of our network.

%
%

The second and third inputs are the left and right context, which will connect to the FOFE encoders. 
Each context is a sequence of sentences $X_C = (s_1,s_2,...s_N)$, where each
sentence is a sequence of word embeddings $s_n = (x_1,x_2,...x_U)$ from the same embedding space as $X$.

%


The first component of the inputs, derived from the focus sentence, is
processed by a bi-directional LSTM with one layer, in order to capture
long-distance dependencies within the sentence. Since LSTMs impose a
significant computational cost for very long sequences we only use this layer
for the input representing the focus sentence, and not for the left and right
contexts.


The LSTM generates outputs $h_{lstm} = (h_1,h_2...,h_T)$ which are passed on to the 
convolutional layer (CNN) in order to learn local features for different kernel sizes $l$
from the history-aware outputs of the LSTM. For each of several kernel sizes, we generate $f$ 
different features, to give CNN outputs $c_{cnn}^l =
(c_1,c_2,\cdots,c_{T-l+1})$. For each CNN output $c_{cnn}^l$, we use max-overtime pooling to extract the most significant feature, and 
dropout to make the learned features more robust.








We use an adapted version of FOFE to provide information about the left and right contexts of the focus. 
Instead of the original 1 of k FOFE representation, we apply FOFE encoding to word2vec embeddings.
This gives a weighted sum of the context word embeddings, with weights
decreasing exponentially with distance from the focus.

The embedding $z$ for a sentence $(x_1,x_2,...x_U)$ is initialised to
$z_1 = x_1$, and then calculated recursively for $u \in 2\cdots U$ as
$z_u = \alpha \cdot z_{u-1} + x_u$.  The parameter $\alpha$ is the forgetting
factor, which controls how fast the weights used for words farther away from
the start of the sentence diminish.  This method is fast and compactly
encodes the words of a sentence in a single embedding vector.



For our use of FOFE, we encode all sentences in the document to left and right
of the focus sentence, in two hierarchical steps. First we encode each context
sentence into a FOFE embedding $z^{sent}$, with a slowly-decreasing
$\alpha_{sent}$. Following this, the left context FOFE encodings are
themselves encoded into a single context embedding using a rapidly decreasing
$\alpha_{cont}$. This is calculated starting with $z^{cont}_1 = z^{sent}_1$
and is calculated for $m \in 2\cdots |C_{left}|$ as
$z^{cont}_m = \alpha_{cont} \cdot z^{cont}_{m-1} + z^{sent}_m$. The right
context FOFE encodings are encoded in the same way, starting with
$z^{cont}_{|C_{right}|} = z^{sent}_{|C_{right}|}$ and recursively applying the
same formula for $m \in |C_{right}|\cdots 2$.  This gives a heavy bias towards
sentences more local to the focus sentence, but only slightly decreases the importance of words
within each sentence. The final FOFE embeddings for the left and right contexts are then put through a dense linear layer to obtain the 
hidden layer outputs, which are combined with the LSTM-CNN outputs. The concatenated outputs from the dense FOFE layers and from the CNN layer for all kernel sizes are 
then used as input to a final softmax output layer.





%
%
%
%

%

\section{Experiments}

We compare the performance of four different network architectures: 1) CNN
only; 2) LSTM only; 3) LSTM-CNN; 4) LSTM context encoded LSTM-CNN
(L-LSTM-CNN), in which the one left and right context sentence are encoded by LSTM; and 
5) Context-LSTM-CNN (C-LSTM-CNN). We use the following two datasets for evaluation:

{\bf Interactive Emotional Dyadic Motion Capture Database \cite{busso2008iemocap}\footnote{\url{http://sail.usc.edu/iemocap/iemocap\_release.htm}} 
(IEMOCAP).}
Originally created for the analysis of human emotions based on speech and video,
a transcript of the speech component is available for NLP research. 
Each sentence in the dialogue is annotated with one of 10 types of emotion. 
There is a class imbalance in the labelled data, and so 
we follow the approach of \cite{chernykh2017emotion}, and only use sentences classified with one of four labels
(`Anger', `Excitement', `Neutral' and `Sadness'). For this dataset, instead of 
using left and right contexts, we assign all sentences from one person to one context and all sentences from the 
other person to the other context. While only the sentences with the four classes of interest are used
for classification, all sentences of the dialog are used as the context. This
results in a set of 4936 labelled sentences with average sentence length 14, and average document length is 986. 

{\bf Drug-related Adverse Effects \cite{gurulingappa2012development}\footnote{\url{https://sites.google.com/site/adecorpus/home/document}} (ADE).} 
This dataset contains sentences sampled from 
the abstracts of medical case reports. For each sentence, the annotation
indicates whether adverse effects of a drug are being described (`Positive') or not (`Negative'). 
The original release of the data does not contain the document context, which
we reconstructed from PubMed\footnote{\url{https://www.ncbi.nlm.nih.gov/pubmed/}}. Sentences 
for which the full abstract could not be found were removed, resulting in
20,040 labelled sentences, with average sentence length 21 and average document length 129.







\begin{table}[!htbp]
\begin{center}
{\small
\begin{tabular}{|l|r|r|r|}
\hline Model      & IEMOCAP & ADE & time(s)\\ \hline
CNN only          & 58.16 {\tiny(0.78)} & 89.49 {\tiny(0.75)} & 218 \\
LSTM only         & 56.30 {\tiny(2.16)} & 89.04 {\tiny(0.75)} & 648 \\
LSTM-CNN          & 59.43 {\tiny(1.60)} & 89.86 {\tiny(1.06)} & 1239 \\
L-LSTM-CNN        & 63.84 {\tiny(2.03)} & 90.22 {\tiny(0.75)} & 1800 \\
C-LSTM-CNN        &{\bf71.49} {\tiny(2.32)} & {\bf90.85} {\tiny(0.37)} & 1243\\
\hline
\end{tabular}
}
\end{center}
\caption{\label{re:TestAccuracy} Average test accuracy and training time. Best values are marked as bold, standard deviations in parentheses}
\end{table}

\begin{table*}[!htbp]
\begin{center}
{\small
\begin{tabular}{|l|r|r|r|r|r|r|}
\hline F-measure     & Anger {\tiny(1,103)} & Sadness {\tiny(1,084)} & Neutral {\tiny(1,708)} & Excitement {\tiny(1,041)} & Negative{\tiny(14,854)} & Positive{\tiny(5,186)} \\ \hline
CNN only          & 67.44 {\tiny(1.02)} & 56.92 {\tiny(3.25)} & 54.93 {\tiny(3.70)} & 53.93 {\tiny(2.50)} & 80.59 {\tiny(1.08)} & 92.28 {\tiny(0.59)} \\
LSTM only         & 65.07 {\tiny(2.49)} & 54.21 {\tiny(4.03)} & 55.12 {\tiny(2.95)} & 49.75 {\tiny(1.80)} & 80.25 {\tiny(1.23)} & 92.24 {\tiny(0.54)}\\
LSTM-CNN          & 67.74 {\tiny(1.11)} & 55.86 {\tiny(6.56)} & 57.17 {\tiny(3.27)} & 56.95 {\tiny(4.06)} & 81.55 {\tiny(0.99)} & 93.00 {\tiny(0.87)}\\
L-LSTM-CNN        & 72.83 {\tiny(1.81)} & 60.35 {\tiny(4.65)} & 61.67 {\tiny(3.18)} & 61.30 {\tiny(2.64)} & 82.29 {\tiny(0.80)} & 93.24 {\tiny(0.60)}\\
C-LSTM-CNN        & {\bf79.54}{\tiny(1.70)} & {\bf66.07}{\tiny(4.65)} & {\bf67.54}{\tiny(4.72)} & {\bf73.11}{\tiny(4.09)} & {\bf83.11}{\tiny(0.24)} & {\bf93.72 }{\tiny(0.33)} \\
\hline
\end{tabular}
}
\end{center}
\caption{\label{re:Fmeasure} Average test F-measure for each class. 
  Instance numbers in parentheses after class name. Best values are marked as bold, standard deviations in parentheses}
\end{table*}

In all experiments, five-fold cross validation was used for evaluation (for comparison with \cite{Huynh2016}). For each fold, 50 epochs were run for 
training using a minibatch size of 64 for each fold, and the Adamax 
optimization algorithm. 
To deal with label imbalance in the data, class weights $w_i$ for class $i$ were set proportional to $\max(f_i)/f_i$ 
where $f_i$ is the frequency of class $i$. 

We used word2vec embeddings with 50 dimensions (suggesed as sufficient by \cite{Lai2016}). For the LSTM, 64 hidden units were used. For the CNN, layers for kernel sizes 2 to
6 were included in the network, and 64 features were used for each.


\subsection{Effect of Forgetting Factors}



\begin{figure}
\includegraphics[width=0.5\textwidth]{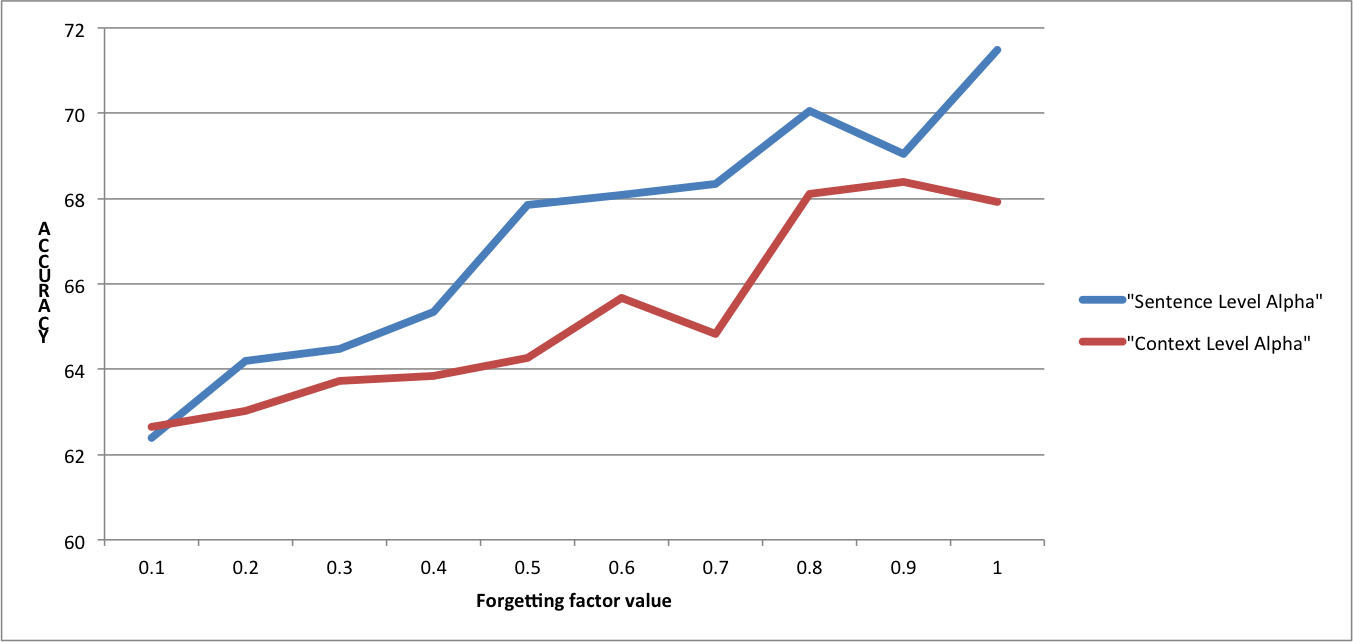}
\caption{Context level (red line) and sentence level (blue line) forgetting factor test}\label{figsa}
\end{figure}

We examined the effect of the two context encoder hyperparameters:
$\alpha_{cont}$ (context level forgetting factor) and $\alpha_{w}$ (sentence level
forgetting factor) on classification performance over the IEMOCAP dataset.  We
tested both in the range of 0.1 to 1 with an incremental step of 0.1. Results
are shown in Figure \ref{figsa}. Accuracy improves as $\alpha_{cont}$ increases,
but drops at $\alpha_{cont} = 1$, at which point all context
sentence are given equal weight. This may be because context closest to the
focus sentence is more important than distant context. Therefore, we select
$\alpha_{cont} = 0.9$ in all experiments.




For $\alpha_{sent}$, performance always increases as $\alpha_{sent}$ increases, with
best results at $\alpha_{sent} = 1$, at which point all words in the sentence
contribute equally in the context code. This implies that for individual sentences in the context, it is more preferable to lose word order, than to down weight any individual word. In all experiments, we therefore set the sentence level forgetting factor to $\alpha_{sent} = 1$


\subsection{Evaluation Results}


Table \ref{re:TestAccuracy} shows the mean and standard deviations for accuracy over the cross validation folds, and training time,
for both data sets. CNN alone performs better than LSTM alone in both tasks. The combined LSTM-CNN network consistently improves performance
beyond both CNN alone and LSTM alone. Both context based models (L-LSTM-CNN
and C-LSTM-CNN) perform better than non context based models, but note that
L-LSTM-CNN increases training time by approximately 1.5x,  whereas
C-LSTM-CNN shows only a marginal increase in training time, with a large
increase in accuracy on the IEMOCAP corpus.

Table \ref{re:Fmeasure} shows the F1-measure for each class in the two datasets. Again, Context-LSTM-CNN outperforms the other models on all classes for all 
data sets. C-LSTM-CNN improves on average by 6.28 over L-LSTM-CNN, 10.16 over LSTM-CNN, 11.4 over CNN and 13.29 over LSTM. 

We conducted a t-test between L-LSTM-CNN and C-LSTM-CNN. On IEMOCAP, C-LSTM-CNN is significantly better than
L-LSTM-CNN ($p = 0.002$). On ADE, C-LSTM-CNN is not significantly
better than L-LSTM-CNN ($p = 0.128$). This may because ADE sentences are less context
dependent. Alternatively, as the ADE task is relatively easy, with all models
able to achieve about 90\% accuracy, a context based approach might not be
able to further improve the accuracy.

\section{Conclusion}
In this paper we introduced a new ANN model, {\bf Context-LSTM-CNN}, that
combines the strength of LSTM and CNN with the lightweight context encoding
algorithm, FOFE.  Our model shows a consistent improvement over either a non-context based model and a LSTM context encoded model, for the sentence classification task.

\section*{Acknowledgements}
This work was partially supported by the European Union
under grant agreement No. 654024 SoBigData.

\bibliography{ref}

\begin{thebibliography}{17}
\expandafter\ifx\csname natexlab\endcsname\relax\def\natexlab#1{#1}\fi

\bibitem[{Busso et~al.(2008)Busso, Bulut, Lee, Kazemzadeh, Mower, Kim, Chang,
  Lee, and Narayanan}]{busso2008iemocap}
Carlos Busso, Murtaza Bulut, Chi-Chun Lee, Abe Kazemzadeh, Emily Mower, Samuel
  Kim, Jeannette~N Chang, Sungbok Lee, and Shrikanth~S Narayanan. 2008.
\newblock Iemocap: Interactive emotional dyadic motion capture database.
\newblock \emph{Language resources and evaluation}, 42(4):335.

\bibitem[{Chernykh et~al.(2017)Chernykh, Sterling, and
  Prihodko}]{chernykh2017emotion}
Vladimir Chernykh, Grigoriy Sterling, and Pavel Prihodko. 2017.
\newblock Emotion recognition from speech with recurrent neural networks.
\newblock \emph{arXiv preprint arXiv:1701.08071}.

\bibitem[{Cho et~al.(2014)Cho, van Merrienboer, Gulcehre, Bougares, Bahdanau,
  Schwenk, and Bengio}]{Cho2014}
Kyunghyun Cho, Bart van Merrienboer, Caglar Gulcehre, Fethi Bougares, Dzmitry
  Bahdanau, Holger Schwenk, and Yoshua Bengio. 2014.
\newblock Learning phrase representations using rnn encoder-decoder for
  statistical machine translation.
\newblock In \emph{The 2014 Conference on Empirical Methods in Natural Language
  Processing}.

\bibitem[{Collobert et~al.(2011)Collobert, Weston, Bottou, Karlen, Kavukcuoglu,
  and Kuksa}]{Collobert2011}
Ronan Collobert, Jason Weston, Léon Bottou, Michael Karlen, Koray Kavukcuoglu,
  and Pavel Kuksa. 2011.
\newblock Natural language processing (almost) from scratch.
\newblock \emph{Journal of Machine Learning Research}.

\bibitem[{Conneau et~al.(2017)Conneau, Schwenk, Barrault, and
  Lecun}]{Conneau2017}
Alexis Conneau, Holger Schwenk, Loïc Barrault, and Yann Lecun. 2017.
\newblock Very deep convolutional networks for text classification.
\newblock In \emph{Proceedings of the 15th Conference of the European Chapter
  of the Association for Computational Linguistics}.

\bibitem[{Gurulingappa et~al.(2012)Gurulingappa, Rajput, Roberts, Fluck,
  Hofmann-Apitius, and Toldo}]{gurulingappa2012development}
Harsha Gurulingappa, Abdul~Mateen Rajput, Angus Roberts, Juliane Fluck, Martin
  Hofmann-Apitius, and Luca Toldo. 2012.
\newblock Development of a benchmark corpus to support the automatic extraction
  of drug-related adverse effects from medical case reports.
\newblock \emph{Journal of biomedical informatics}, 45(5):885--892.

\bibitem[{Hochreiter and Schmidhuber(1997)}]{Hochreiter1997}
Sepp Hochreiter and Jürgen Schmidhuber. 1997.
\newblock Long short-term memory.
\newblock \emph{Neural Computation}.

\bibitem[{Huynh et~al.(2016)Huynh, He, Willis, and Rueger}]{Huynh2016}
Trung Huynh, Yulan He, Allistair Willis, and Stefan Rueger. 2016.
\newblock Adverse drug reaction classification with deep neural networks.
\newblock In \emph{Proceedings of The 26th International Conference on
  Computational Linguistics}.

\bibitem[{Kim(2014)}]{Kim2014}
Yoon Kim. 2014.
\newblock Convolutional neural networks for sentence classification.
\newblock In \emph{The 2014 Conference on Empirical Methods in Natural Language
  Processing}.

\bibitem[{Kingma and Ba(2015)}]{Kingma2015}
Diederik~P. Kingma and Jimmy Ba. 2015.
\newblock Adam: A method for stochastic optimization.
\newblock In \emph{The 3rd International Conference for Learning
  Representations}.

\bibitem[{Lee and Dernoncourt(2016)}]{Lee2016}
Ji~Young Lee and Franck Dernoncourt. 2016.
\newblock Sequential short-text classification with recurrent and convolutional
  neural networks.
\newblock \emph{Proceedings of the 2016 Conference of the North American
  Chapter of the Association for Computational Linguistics: Human Language
  Technologies}.

\bibitem[{Mikolov et~al.(2013{\natexlab{a}})Mikolov, Chen, Corrado, and
  Dean}]{mikolov2013efficient}
Tomas Mikolov, Kai Chen, Greg Corrado, and Jeffrey Dean. 2013{\natexlab{a}}.
\newblock Efficient estimation of word representations in vector space.
\newblock \emph{arXiv preprint arXiv:1301.3781}.

\bibitem[{Mikolov et~al.(2013{\natexlab{b}})Mikolov, Sutskever, Chen, Corrado,
  and Dean}]{NIPS2013_5021}
Tomas Mikolov, Ilya Sutskever, Kai Chen, Greg~S Corrado, and Jeff Dean.
  2013{\natexlab{b}}.
\newblock Distributed representations of words and phrases and their
  compositionality.
\newblock In C.~J.~C. Burges, L.~Bottou, M.~Welling, Z.~Ghahramani, and K.~Q.
  Weinberger, editors, \emph{Advances in Neural Information Processing Systems
  26}, pages 3111--3119. Curran Associates, Inc.

\bibitem[{Sainath et~al.(2015)Sainath, Vinyals, Senior, and Sak}]{Sainath2015}
Tara Sainath, Oriol Vinyals, Andrew Senior, and Hasim Sak. 2015.
\newblock Convolutional, long short-term memory, fully connected deep neural
  networks.
\newblock In \emph{IEEE International Conference on Acoustics, Speech and
  Signal Processing}.

\bibitem[{Xu et~al.(2015)Xu, Ba, Kiros, Cho, Courville, Salakhutdinov, Zemel,
  and Bengio}]{Xu2015}
Kelvin Xu, Jimmy Ba, Ryan Kiros, Kyunghyun Cho, Aaron Courville, Ruslan
  Salakhutdinov, Richard Zemel, and Yoshua Bengio. 2015.
\newblock Show, attend and tell: Neural image caption generation with visual
  attention.
\newblock \emph{Proceedings of 2015th International Conference on Machine
  Learning}.

\bibitem[{Zhang et~al.(2015)Zhang, Jiang, Xu, Hou, and Dai}]{Zhang2015}
Shiliang Zhang, Hui Jiang, Mingbin Xu, Junfeng Hou, and Lirong Dai. 2015.
\newblock The fixed-size ordinally-forgetting encoding method for neural
  network language models.
\newblock In \emph{Proceedings of the 53rd Annual Meeting of the Association
  for Computational Linguistics and the 7th International Joint Conference on
  Natural Language Processing}.

\bibitem[{Zhou et~al.(2015)Zhou, Sun, Liu, and Lau}]{Zhou2015}
Chunting Zhou, Chonglin Sun, Zhiyuan Liu, and Francis C.~M. Lau. 2015.
\newblock A c-lstm neural network for text classification.
\newblock \emph{arXiv151108630Z}.

\end{thebibliography}
\bibliographystyle{acl_natbib_nourl}

\end{document}